\documentclass[prb,twocolumn,letter]{revtex4}
\usepackage{amssymb}
\usepackage{amsmath}
\usepackage{bbm}
\usepackage{graphicx}

\begin{document}

\title{Doping-driven Mott transition in the one-band Hubbard model}
\author{Philipp Werner}
\author{Andrew J. Millis}
\affiliation{Columbia University, 538 West, 120th Street, New York, NY 10027, USA}

\date{October 14, 2006}

\hyphenation{}

\begin{abstract}
A powerful new impurity solver is shown to permit  a systematic study of the 
doping driven Mott transition in a one-band Hubbard model within the framework 
of single-site dynamical mean field theory. At small dopings and large interaction strengths
we are able access low enough  temperatures that a reliable extrapolation to 
temperature $T=0$ may be performed, and ground state energies of insulating
and metallic states may be compared. We find that the $T=0$ doping-driven  transition is 
of second order and is characterized by an interaction-strength dependent electronic compressibility, which vanishes at the critical interaction strength of the half filled model.
Over wide parameter ranges the compressibility is substantially reduced relative to the 
non-interacting system. The metal insulator transition is characterized by the appearance of in-gap
states, but these are relevant only for very low dopings of less than $3\%$.
\end{abstract}


\maketitle

\section{Introduction}

The `Mott' or correlation-induced insulating state is a fundamental unifying concept in 
modern condensed matter physics. The physical properties of many interesting materials,
including organic conductors,\cite{organicref} colossal magnetoresistance manganites,\cite{CMRMott}
actinides such as Ce and Pu,\cite{Ceref,Puref} and many transition metal oxide compounds
\cite{Imada98} are believed to be controlled by proximity to a Mott insulating state.  However, the accurate
theoretical description of the physics of Mott insulators poses challenging problems, and many 
questions remain unresolved.   A particularly important class of open questions, crucial,
for example to the physics of high temperature superconductivity,\cite{Hightcref} concerns
the behavior at strong interactions as the carrier concentration is varied away from the commensurate values
at which Mott insulating behavior occurs.  

\begin{figure}[h!]
\centering
\includegraphics [angle=0, width= 8cm] {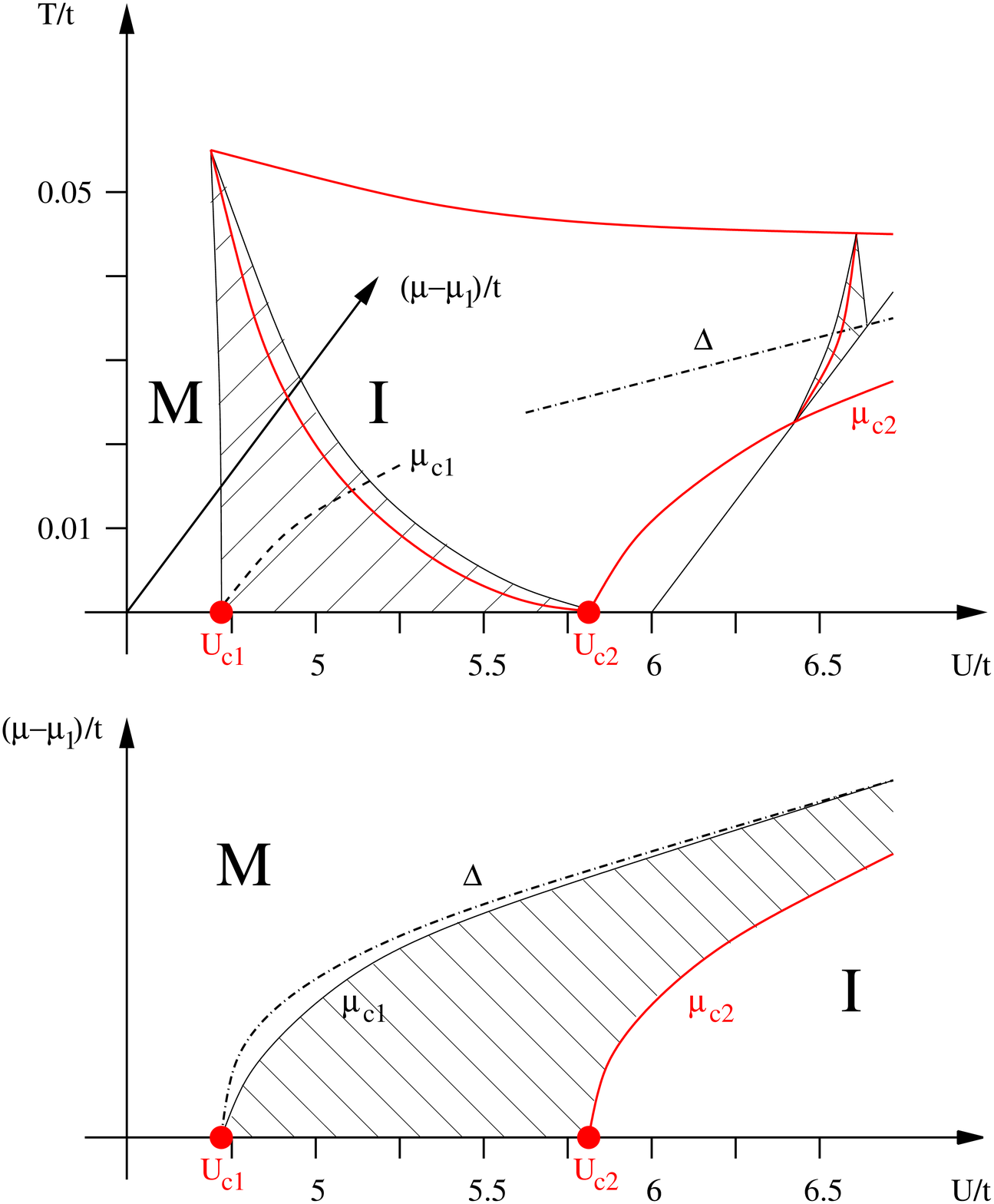}
\caption{(color online) Sketch of the phase diagram of the Hubbard model in the single site DMFT approximation (semi-circular density of states of bandwidth $4t$) with magnetic order suppressed by averaging over spin. Top panel: thick red lines indicate the surface in the space of temperature, interaction strength and chemical potential, where a first order metal insulator transition occurs. Thin black lines delimit the coexistence region (hashed for cuts across the $\mu=\mu_1$ and $U=6t$ planes) where both metallic and insulating solutions to the DMFT equations exist. Lower panel: $T=0$ phase diagram and spectroscopic gap $\Delta$.   
}
\label{phasediagram}
\end{figure}

An important theoretical step forward was achieved with the development of 
``dynamical mean field theory",\cite{Georges96} which showed that if the momentum 
($p$) dependence of the electron self energy $\Sigma(p,\omega)$ can be appropriately 
approximated,   the computation of electronic properties may be reduced to the solution
of a quantum impurity model, along with a self consistency condition. In
``single-site"  dynamical mean field theory, the momentum dependence is completely 
neglected, $\Sigma(p,\omega)\rightarrow \Sigma(\omega)$, and the impurity model is a 
single site coupled to a free fermion bath. 
This approximation is strictly valid in the limit of inifinite
coordination number, but captures many features of the behavior of finite dimensional compounds.\cite{organicref,Ceref,Puref,CMRMott}

One of the  early successes of the single-site dynamical mean field theory was an explication of the 
basic phase diagram of the
Mott transition in single band (Hubbard-like) models.\cite{KotliarUc2} 
If magnetism may be neglected, the phase diagram in the space spanned by temperature ($T$), interaction strength ($U$) and chemical potential ($\mu$) takes the form shown in the upper panel of Fig.~\ref{phasediagram}. At low, but non-zero temperature, a first order metal-insulator transition occurs on the surface delimited by thick red lines. The coexistence region, where both metallic and insulating solutions to the DMFT equations exist, is indicated by hashed areas on the planes $\mu=\mu_1(U)$ (corresponding to half-filling) and $U=6t$. The $T=0$
phase diagram is shown in the lower panel of Fig.~\ref{phasediagram}.  If the chemical potential is held at the value $\mu_1(U)$, then the model is metallic for interactions weaker than a critical value 
conventionally denoted $U_{c2}$ and we expect properties varying smoothly with chemical potential. 
As $U\rightarrow U_{c2}$ from below, the quasiparticle weight and compressibility vanish continuously, so that the metal-insulator transition at $T=0$ is second order. 
For $U>U_{c2}$ the model is insulating for carrier concentration $n=0.5$ per spin, and there is correspondingly a region of chemical potential (bounded by the curve labeled $\mu_{c2}$), where the density is pinned at $2n=1$ and no metallic solution exists.  

The second order character of the transition at $U_{c2}$ strongly suggests that the
critical chemical potential $\mu_{c2}$ smoothly approaches $\mu_1$ as $U \rightarrow U_{c2}$ from above.
However, the insulating phase at $U>U_{c2}$ is characterized by a spectroscopic gap $\Delta$
which does not vanish at $U=U_{c2}^+$
and remains locally stable 
for a range of $U<U_{c2}$ (the lower limit of the region 
in which the insulating solution exists, is conventionally denoted by $U_{c1}$).  
The gap, presented as a chemical potential difference
from the half filled value $\mu_1$ is shown as a dashed-dotted line in Fig.~\ref{phasediagram}.
The difference between the values of $\Delta$ and $\mu_{c2}$ implies that in this approximation, doping
a Mott insulator produces ``in-gap" states, and Fisher, Kotliar and Moeller have presented more precise
analytical arguments which support these ideas and show that 
doping generates in-gap states  for all $U>U_{c2}$.\cite{Fisher95, Kajueter95}

However, moving beyond  general arguments, very little is known with confidence about the specifics of 
the phase diagram. The essential difficulty has been the lack of numerical methods powerful 
enough to address the region of strong correlations, low temperature and low doping 
where the interesting physics occurs.  
In this paper we use a newly developed method \cite{Werner05,Werner06} to solve the problem. 
The method permits access to strong correlations and low temperatures, with an unprecedented
accuracy which enables us to construct thermodynamic potential curves and establish
the nature and location of the transition. 
Our results are consistent with the following scenario: the point $U=U_{c2}$ and $\mu=\mu_1$ is a 
quantum critical point at which the electronic compressibility (proportional to $\partial n/\partial \mu$) vanishes
linearly in $|\mu-\mu_1|$. For $U>U_{c2}$ and $T=0$, a second order metal-insulator transition 
occurs at an interaction dependent chemical potential $\mu_{c2}(U)-\mu_1\sim (U-U_{c2})^x$, 
with an exponent $x$ close to 
$1/2$. For $U>U_{c2}$ the compressibility does not vanish as $\mu\rightarrow \mu_{c2}$, but
at large enough $\mu-\mu_1$, $\partial n/\partial \mu \sim |\mu-\mu_1|$. The physics associated with the 
critical point is visible over a reasonable range of $U>U_{c2}$.

\section{Model and Formalism}
\label{method}

\subsection{Model}

In this paper we present results for the paradigm strongly correlated model, the one-orbital Hubbard
model, defined on a lattice of sites $i$ by
\begin{equation}
H=-\sum_{i,\delta,\sigma}t(\delta)d^\dagger_{i+\delta,\sigma}d_{i,\sigma}+\sum_iUn_{i,\uparrow}n_{i,\downarrow}.
\label{HHub}
\end{equation}
The energy dispersion $\varepsilon_p$ is defined as the Fourier transform of $t(\delta)$ and the
only property of the dispersion which will be important for us is 
the density of states ${\cal D}(\omega)=\sum_p \delta(\omega-\varepsilon_p)$. In our 
specific calculations we shall take 
\begin{equation}
{\cal D}(\omega)=\sqrt{4t^2-\omega^2}/(2\pi t^2).
\label{dos}
\end{equation}

For this choice of density of states
the chemical potential $\mu$ corresponding
to the (potentially Mott insulating) density of one electron per site is $\mu_1=U/2$
and the critical interaction strength for the zero temperature 
Mott transition in the single-site DMFT approximation
is $U_{c2} \approx 5.8t$.\cite{Bulla99}

\subsection{DMFT method}

The single-site DMFT reduces the solution of the lattice problem to the solution of
a quantum impurity problem defined by
\begin{eqnarray}
H_{QI}&=&-\sum_\sigma \mu d^\dagger_{\sigma}d^{\phantom\dagger}_\sigma +Un_{d,\uparrow}n_{d,\downarrow}
\nonumber \\
&&+\sum_{\varepsilon,\sigma}
\left(V_\varepsilon d^\dagger_\sigma c_{\varepsilon,\sigma}+V^*_\varepsilon c^\dagger_{\varepsilon,\sigma} d_\sigma 
+\varepsilon c^\dagger_{\varepsilon,\sigma}c^{\phantom\dagger}_{\varepsilon,\sigma} \right).
\label{HQI}
\end{eqnarray}
It is useful to define the hybridization function \cite{Werner06}
\begin{equation}
F_\sigma(\tau)=\sum_\varepsilon \left|V_\varepsilon\right|^2\langle T_\tau c^\dagger_{\varepsilon,\sigma}(\tau)c_{\varepsilon,\sigma}(0) \rangle_\text{bath},
\label{F}
\end{equation}
the $d$-electron Green function $G_\sigma(\tau)\equiv -\left<T_\tau d_\sigma(\tau) d^\dagger_\sigma(0)\right>$
and self energy $\Sigma_\sigma\equiv \partial_\tau+\mu+F_\sigma(\beta-\tau)-G_\sigma(\tau)^{-1}$.  The hybridization
function is fixed by the self consistency condition 
\begin{equation}
G_\sigma(\omega)=\int d\varepsilon \frac{ {\cal D}(\varepsilon)}{\omega-\varepsilon-\mu-\Sigma_\sigma(\omega)}.
\label{sce}
\end{equation}

The challenging numerical task is computing $G(\tau)$.  In order to do this we have recently
developed a new  solver,\cite{Werner05, Werner06} which is 
based on a diagrammatic expansion of the partition function in the impurity-bath 
hybridizations and the Monte Carlo sampling of certain collections of the resulting diagrams. 
The summation of diagrams into determinants eliminates the sign problem, even away from half-filling, and our approach, which expands around an exactly solved atomic limit, leads to lower perturbation orders at stronger interactions $U$.  The method thus allows unprecedented access to low temperatures and strong interactions\cite{Gull06} and will be used here
to study the doping-dependent Mott transition. 
Near the end points $\tau=0$ and $\beta$ the Green function converges very rapidly, but 
more effort is needed to accurately determine the long-time behavior. We chose a 
resolution of 10000 points for the Green function and a smoothing procedure 
(averaging over 30 neighboring bins) at intermediate $\tau$ to reduce the 
statistical errors in the region where such a high resolution is not necessary. 
With this resolution, the systematic errors should be small and we therefore 
estimated the error bars on quantities such as densities and energies from 
their variation in successive iterations of the converged solution. Where no error 
bars are given, the errors are smaller than the symbol size.  

From the computed $G(\tau)$ we directly obtain the density per spin $n=G(\tau\rightarrow \beta)$, 
while the ``internal energy" may be computed as
\begin{eqnarray}
E(\mu, T) &=& \left<H-\mu N\right>
\nonumber\\
&=& 2t^2 \int_0^\beta d\tau G(\tau)G(-\tau) + U D - 2\mu n.
\label{energy}
\end{eqnarray}
Here we have used a property of the semicircular density of states to obtain
a compact expression for the kinetic energy term, while the expectation
value of the interaction term is obtained from a direct measurement of the double occupancy $D$.

\subsection{Extrapolation to $T=0$}

As will be shown below, a characterization of the metal-insulator transition
requires the construction of thermodynamic potentials and the extrapolation of our data
to $T=0$. The insulating state is characterized by a large gap
(which means that the energy at low temperature is exponentially close to the ground
state value) and an extensive spin entropy of $\ln 2$ per site,
so the thermodynamic potential of the insulating state is
\begin{equation}
\Omega_\text{ins}=E_\text{ins}-T\ln(2), 
\label{Fins}
\end{equation}
where $E_\text{ins}$ is computed from Eq.~(\ref{energy}).

The entropy of the metallic state is in general not easy to obtain.
However we note that within single-site dynamical mean field theory the metallic
phase is, at low temperatures, a Fermi liquid characterized by a $T^2$ variation of
physical quantities. In particular, at sufficiently low $T$, the energy and thermodynamic potential of the metallic state are given respectively by
\begin{eqnarray}
E_\text{met}(\mu,T)=E_\text{met}(\mu,T=0)+\frac{1}{2}\gamma T^2, 
\label{EofT} \\
\Omega_\text{met}(\mu,T)=E_\text{met}(\mu,T=0)-\frac{1}{2}\gamma T^2, 
\label{FofT} 
\end{eqnarray}
where the specific heat coefficient $\gamma \equiv \lim_{T\rightarrow 0}C/T$ is %given by 
\begin{equation}
\gamma=\frac{2\pi^2}{3}{\cal D}(\mu_0)\left(1-\left.\frac{\partial \Sigma}{\partial \omega}\right|_{\omega=0}\right)\approx \frac{2}{t}\left(1-\frac{\Im m\Sigma(i\omega_0)}{\omega_0}\right).
\label{C_sigma}
\end{equation}
Here $\mu_0$ is the chemical potential which, in the model with $U=0$, produces the density corresponding to $\mu$ in the interacting model. In the last line we used the fact that for the dopings of interest the density of states may be approximated by its half-filled value and that the 
self energy derivative at low $T$ may be approximated by the value of the imaginary part of the self energy at the lowest Matsubara frequency divided by $\pi T$.

We estimate $\gamma$ in two ways: from the electron self energy, via Eq.~(\ref{C_sigma}), or by fitting
the measured energies to a $T^2$ dependence and using Eq.~(\ref{EofT}). 
For $\mu \rightarrow \mu_{c2}$
(especially near $U_{c2}$) the range over which the energy obeys a $T^2$ law becomes small and we find that obtaining $\gamma$ from the self energy leads to smaller errors, which we estimate to be at the
$10\%$  level. We have verified that for the dopings considered, we can reach low enough temperatures that  the metallic entropy $S=\gamma T$ is much smaller than $\ln 2$.

Finally, we note that the thermodynamic potential may alternatively be obtained from the density-chemical potential trace
via the thermodynamic relation $2n=-\partial \Omega/\partial \mu$. Choosing a reference chemical potential 
$\mu_c$  we have
\begin{equation}
\Omega_\text{met}(\mu)=\Omega_\text{met}(\mu_c)-2\int_{\mu_c}^{\mu}d\mu'n(\mu').
\label{Ffromn}
\end{equation}

Below, we will use the $T=0$ limit of Eq.~(\ref{Ffromn}) to demonstrate the consistency
of our analysis and show that at $T=0$ the doping driven transition
becomes second order.

\begin{figure}[t]
\centering
\includegraphics [angle=-90, width= 8.5cm] {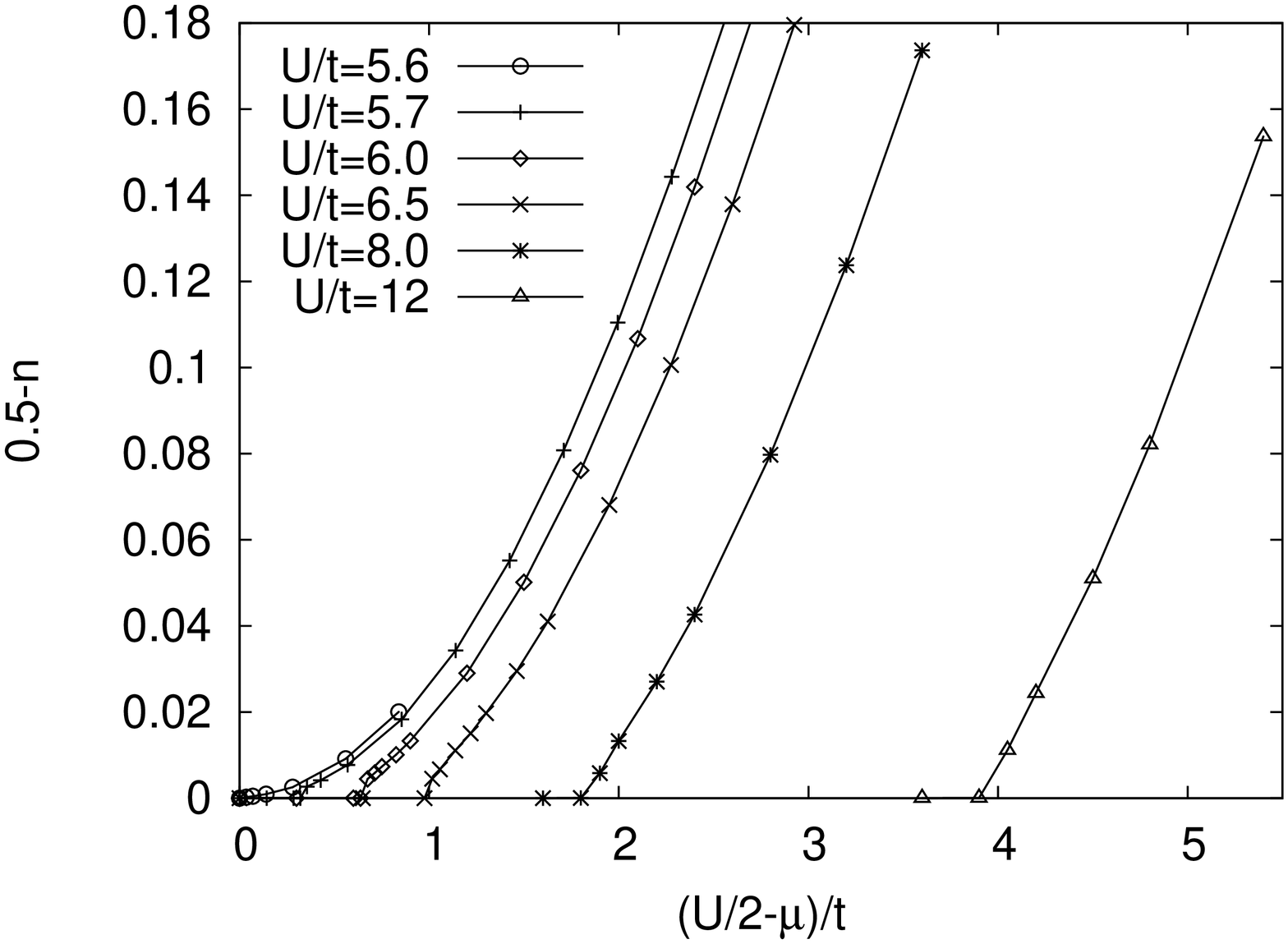}
\includegraphics [angle=-90, width= 8.5cm] {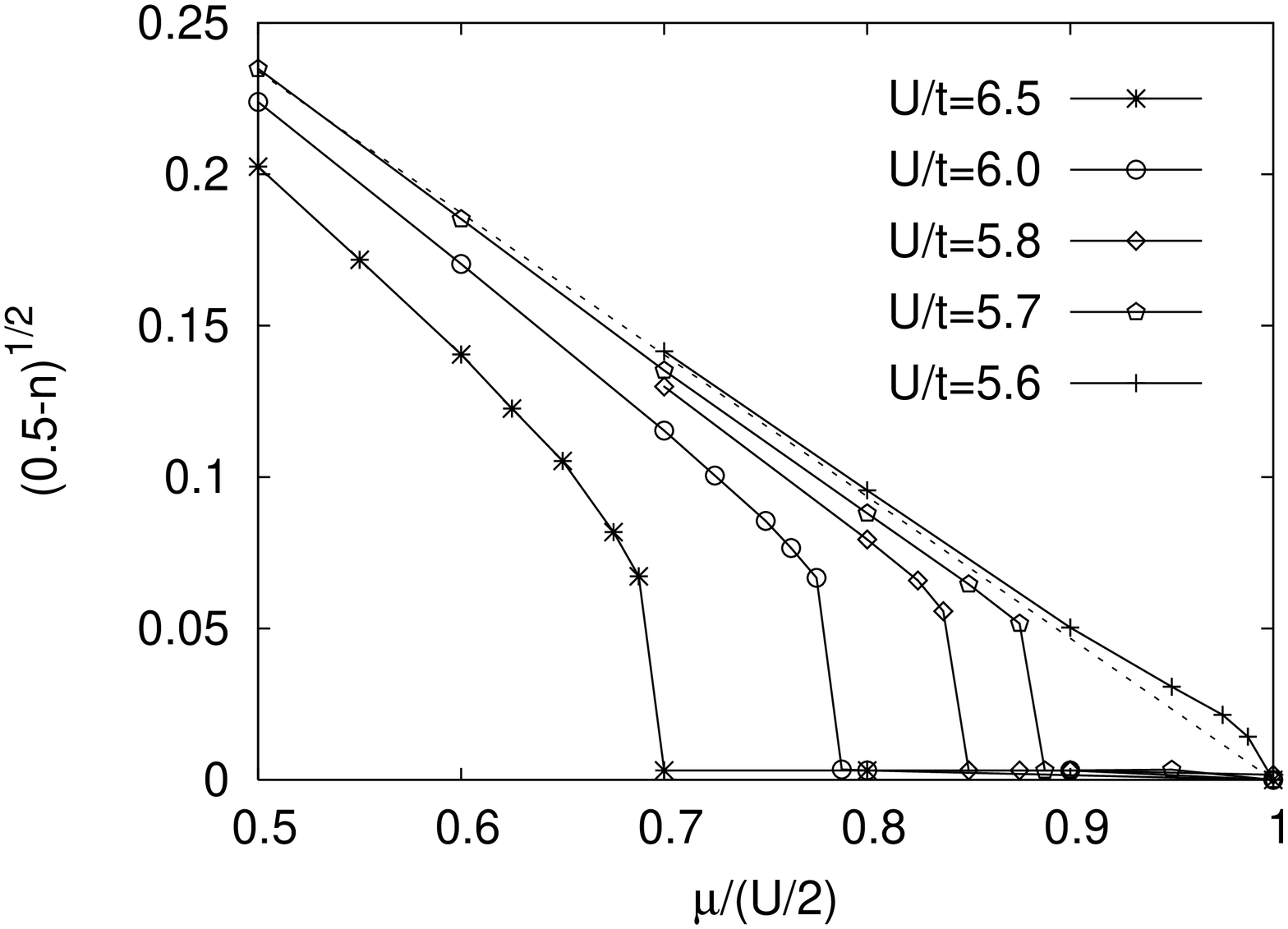}
\caption{
Doping per spin, $0.5-n$,  as a function of chemical potential for $\beta t=400$ and indicated values of $U/t$.  At this temperature, the transition at half-filling ($\mu=\mu_1=U/2$) occurs at $U_{c}(T)\approx 5.65$. 
For larger interactions, a gap opens and shifting the chemical potential 
induces a first order metal-insulator transition. 
Near the critical point $U_{c2}$, $n-0.5\sim (\mu-\mu_1)^2$, but as one moves away from the critical point, the onset of doping becomes linear in $\mu$.
}
\label{n_mu_t}
\label{n}
\end{figure}

\section{Results}

\subsection{First order transition at $T>0$}

The upper panel of Fig.~\ref{n} shows the variation with chemical potential, $\mu$,
of the carrier concentration per spin, $n$, measured relative to the 
Mott insulating value $0.5$ at the very low temperature $t/T\equiv \beta t=400$. 
For the weakest interaction
strength, $U=5.6t$, the carrier concentration varies smoothly with chemical potential,
implying that the phase is metallic even at half filling. For the larger interaction strengths
a gap (region where $n$ is approximately independent of $\mu$) is visible, showing
that for these $U$-values the model at $\beta t=400$ is insulating for a range of chemical potentials. 
At this temperature the critical interaction strength for the Mott transition 
is therefore $U/t\approx 5.65$, consistent with the
accepted phase diagram of the half filled model.\cite{BluemerPhD} 

The density-chemical potential traces shown in the lower panel of Fig.~\ref{n}   
highlight an unusual scaling behavior
near the critical point at $U=U_{c2}$ and $n=0.5$ ($\mu=\mu_1=U/2$): 
the density (measured from $0.5$) varies as the square of the chemical potential (measured
from $\mu_1$), in other words, near the Mott point, the compressibility per spin, which is up to a factor $1/(2n)^2$ given by $\partial n/\partial \mu$, 
vanishes proportionally to  $|\mu-\mu_1|$ with a coefficient $0.44/{\mu_{1}}^2$. 
The $U=5.6t$ curve exhibits at the smallest $\mu$ a crossover away from the square root behavior to the
constant $\partial n/\partial \mu$ expected in a metallic phase.

For $U$ larger than the critical value, the curves exhibit a slight downward
trend away from the  $(\mu-\mu_1)^2$ scaling, indicating a linear onset at very small dopings, 
but more importantly the curves are cut off by a discontinuity indicating our inability to numerically stabilize
a metallic phase, and suggesting 
that at $T>0$ the doping driven metal-insulator transition is first order. The first order transition also
occurs in the interaction driven ($n=0.5$) case \cite{Georges96} and is expected from the 
extensive entropy ($\ln 2$ per site) 
of the paramagnetic insulating state.

\begin{figure}[t]
\centering
\includegraphics [angle=-90, width= 8.5cm] {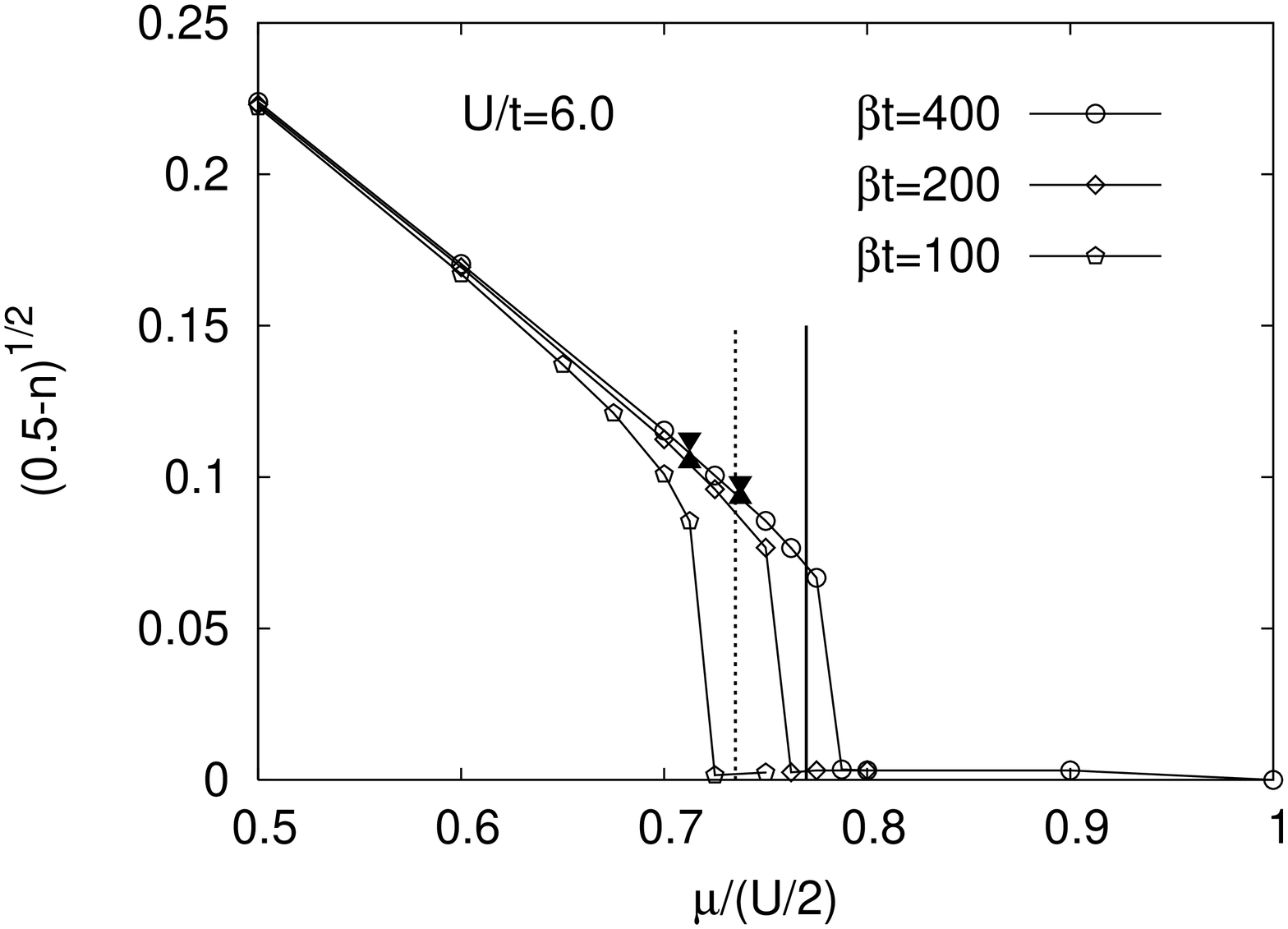}
\includegraphics [angle=-90, width= 8.5cm] {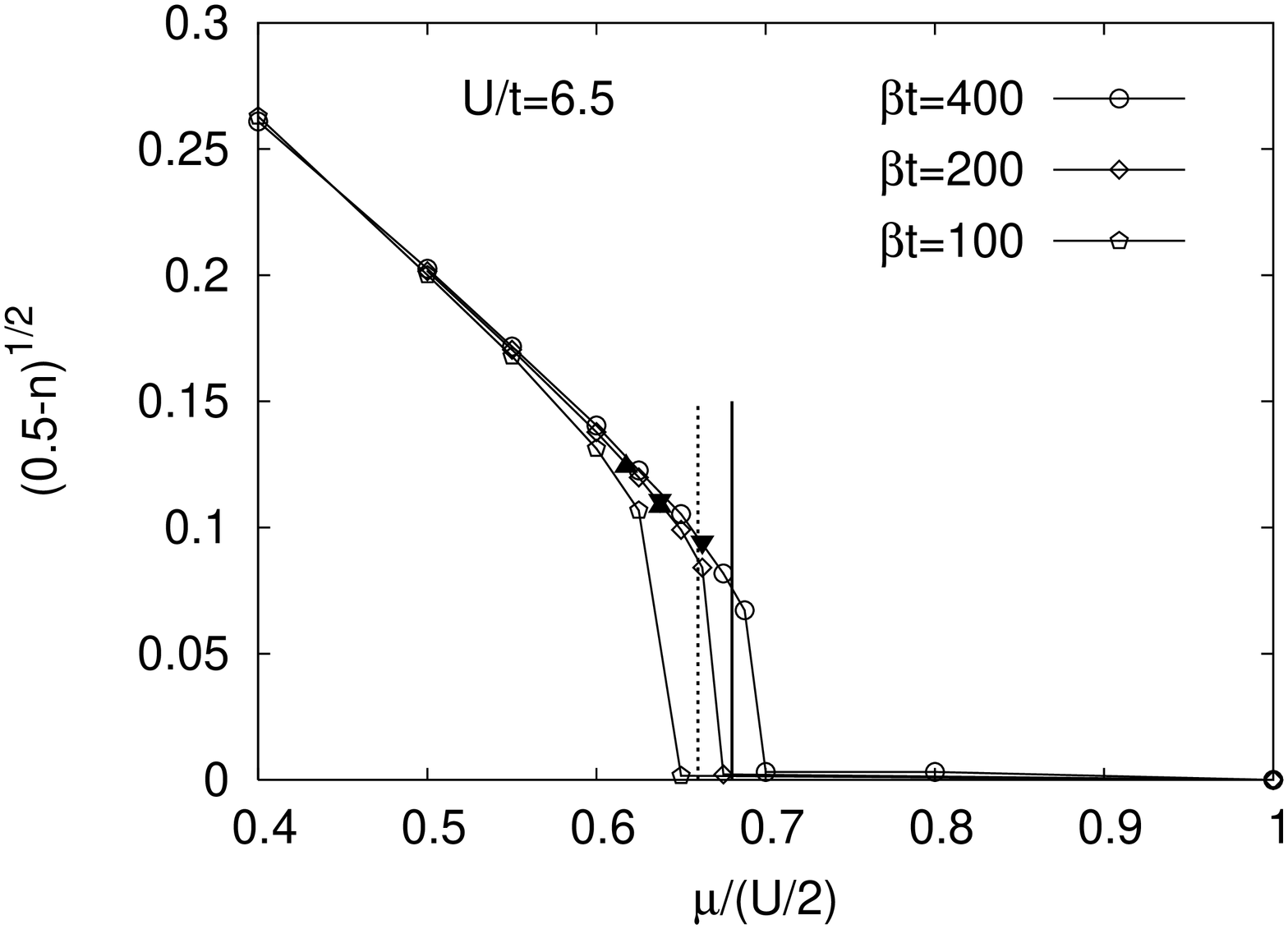}
\caption{Doping-vs-$\mu$ for $U/t=6$ (upper panel), $6.5$ (lower panel) and $\beta t=100$, 200 and 400. For fixed $\mu$, the doping increases with 
decreasing temperature. The vertical lines indicate the positions of the first order phase transition, 
as estimated from the crossing of the thermodynamic potential curves (dashed: $\beta t=200$, solid: $\beta t=400$).
Solid triangles show the densities obtained from the thermodynamic potential using $2n=-\partial \Omega/\partial \mu$ (up-triangles: $\beta t=200$, down-triangles: $\beta t=400$).
}
\label{n_u}
%\label{n_u65}
\end{figure}

We now consider in more detail the behavior at very small dopings. The two panels of Fig.~\ref{n_u} show density-chemical potential traces for three low temperatures $T/t=0.01$, $0.005$ and $0.0025$. 
These results show that for dopings (per spin) $\gtrsim 0.01$, the lowest temperature data are 
essentially converged to the $T=0$ result, while at the smallest doping some temperature dependence
clearly remains.

\begin{figure}[t]
\centering
\includegraphics [angle=-90, width= 8.5cm] {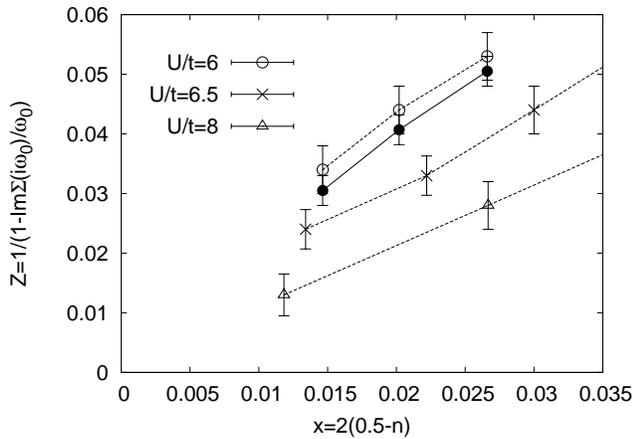}
\caption{Quasi-particle weights estimated using $Z\approx 1/(1-\Im m \Sigma(\omega_0)/\omega_0)$, 
evaluated at the lowest Matsubara frequency $\omega_0$,
plotted as a function of doping per spin. The data points connected by dashed lines correspond to $\beta t=400$ and approximation (\ref{C_sigma}). Solid dots show an estimate for $Z$ which is based on a 2-parameter fitting function.
%$(\mu/(U/2), \gamma)=(0.7, 38),(0,725,  45),(0.75, 59)$ for $U/t=6$ and 
%$(\mu/(U/2), \gamma)=(0.6, 33),(0.625,  43),(0.65, 61)$ for $U/t=6.5$.
}
\label{Z_doping}
\end{figure}

The data presented in Fig.~\ref{n_u} 
define the range of parameters over which the metallic state can be stabilized by our numerical procedure. 
The jump in density suggests the presence, in the $T>0$ phase diagram, of a first order metal-insulator
transition, but the computed jump position is a spinodal point.
To locate the first order phase transition, 
we compute the thermodynamic potentials $\Omega_\text{ins}$ and $\Omega_\text{met}$ 
using Eqs.~(\ref{Fins}) and (\ref{FofT}) and the specific heat coefficients obtained from the self energies. Representative examples 
are shown in Fig.~\ref{Z_doping}
which plots an approximation to the quasiparticle weight $Z=1/(1-\partial \Sigma/\partial \omega | _{\omega\rightarrow 0})\approx 1/(1-\Im m \Sigma(\omega_0)/\omega_0)$ as a function of doping for $U/t=6, 6.5$ and $8$ and several 
chemical potentials. Note that if we define total doping $x=2(0.5-n)$, then our data for $U/t=8$ 
are roughly consistent with $\gamma t=1.9/x$, those at  $U/t=6.5$  with $\gamma t=1.35/x$ and those at $U/t=6$ with $\gamma t= 1.0/x$. 
Especially near $U_{c2}$, the measured $Z$ do not quite extrapolate to zero as $x\rightarrow 0$, which may be due to the approximation of the derivative in Eq.~(\ref{C_sigma}). The full circles in Fig.~\ref{Z_doping} show estimates of $Z$ for $U/t=6$, which were obtained by fitting the $y=\Im m\Sigma(\omega_n)$ data for the lowest three Matsubara frequencies $x=\omega_n$ to a function of the form $x=A y + B y^2$. While a careful examination of the behavior of $Z$ near $U=U_{c2}$ may be desirable, these uncertainties do not affect the analysis in this paper. Using the approximation %$\partial \Sigma/\partial \omega | _{\omega\rightarrow 0}\approx \Im m \Sigma(\omega_0)/\omega_0$, 
in Eq.~(\ref{C_sigma}), the $\beta t=400$ data yield the specific heat coefficients $(\mu/(U/2), \gamma t)=(0.7, 38),(0.725,  45),(0.75, 59)$ for $U/t=6$ and $(0.6, 33),(0.625,  43),(0.65, 61)$ for $U/t=6.5$.

\begin{figure}[t]
\centering
\includegraphics [angle=-90, width= 8.5cm] {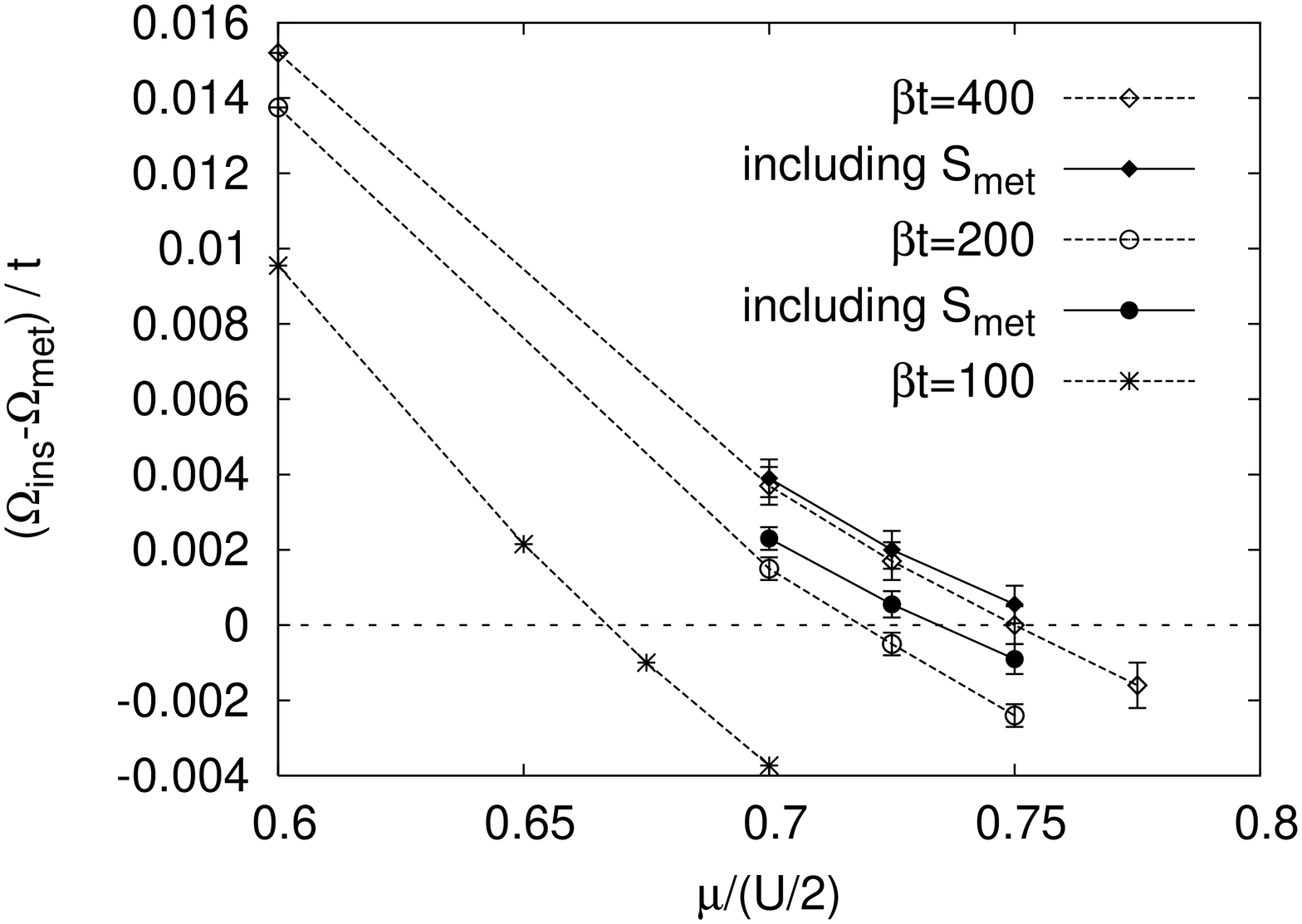}
\includegraphics [angle=-90, width= 8.5cm] {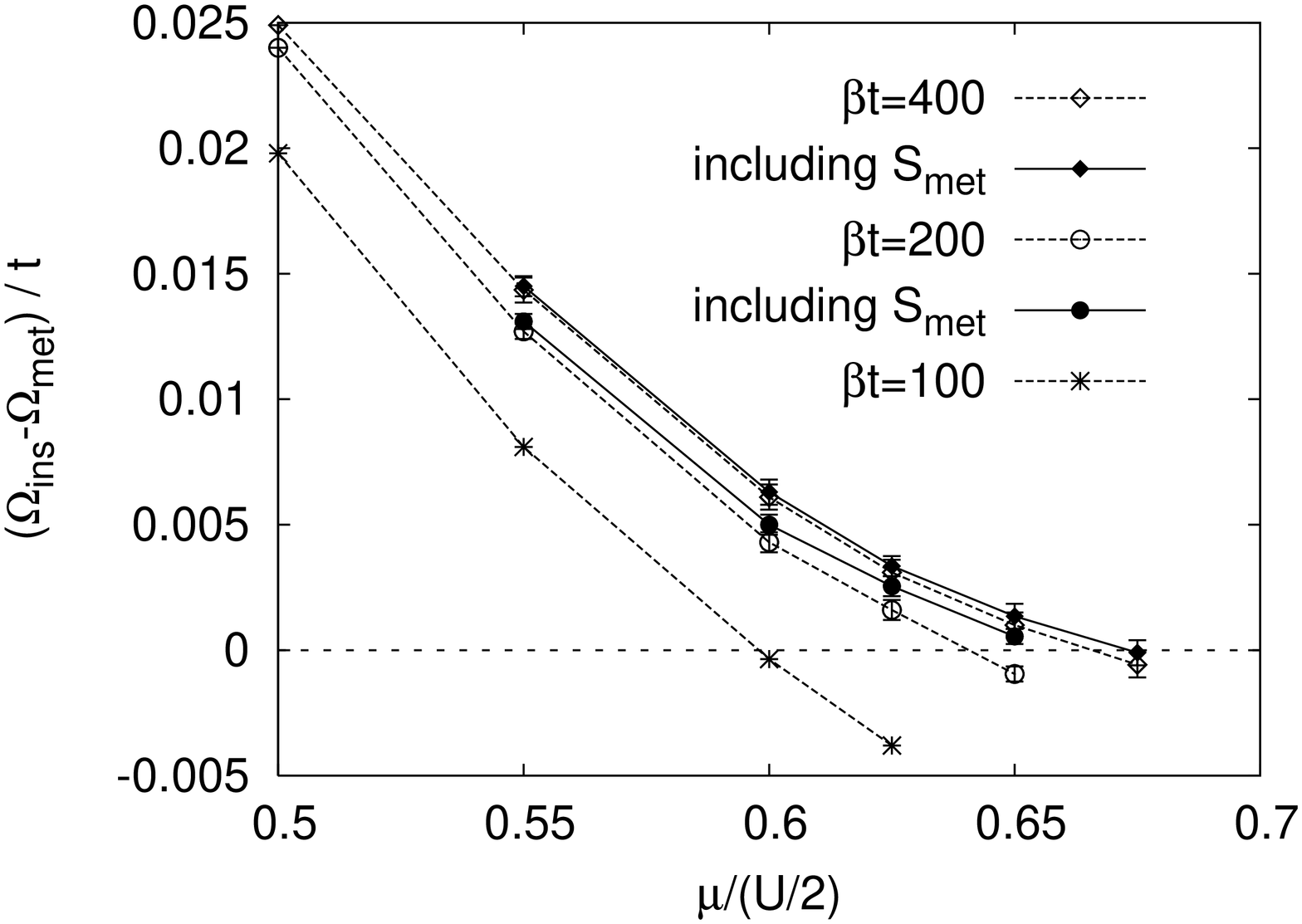}
\caption{Solid points: thermodynamic potential difference between the metallic and insulating solutions for 
the indicated temperatures, $U=6t$ (upper panel) and $U=6.5t$ (lower panel). 
Open symbols: ``thermodynamic potential" computed by neglecting the $1/2\gamma T^2$ term
in Eq.~(\ref{FofT}).
}
\label{Fdiff}
\end{figure}

Figure~\ref{Fdiff} shows the thermodynamic potential differences between metallic and insulating solutions
as a function of chemical potential for $U/t=6$ (upper panel) and $U/t=6.5$ (lower panel).
Because the thermodynamic potential differences are very tiny, taking proper account of the entropy
of the metallic state is important.  
The point where the curve crosses zero yields the location of the first order transition, 
which is indicated by the vertical lines in Fig.~\ref{n_u}. 
The phase transition occurs near the spinodal point and shifts with temperature in a similar way as the spinodal point.

As a consistency check, we show the value of the doping obtained from the thermodynamic potential using the formula
\begin{equation}
2 n_{\text{met}}=-\frac{\partial \Omega_\text{met}}{\partial \mu}
\label{n_F}
\end{equation}
as triangles in Fig.~\ref{n_u}. These results, based on an approximation of the derivative by
the slopes of the solid lines in Fig.~\ref{Fdiff}, agree within 10\% with the measured dopings. The thermodynamic potential curves at larger $|\mu-\mu_1|$ show similar slopes for $\beta t=100$, 200 and 400, and thus yield similar dopings. So, within the expected precision our thermodynamic potential analysis yields consistent results.

\begin{figure}[ht]
\centering
\includegraphics [angle=-90, width= 8.5cm] {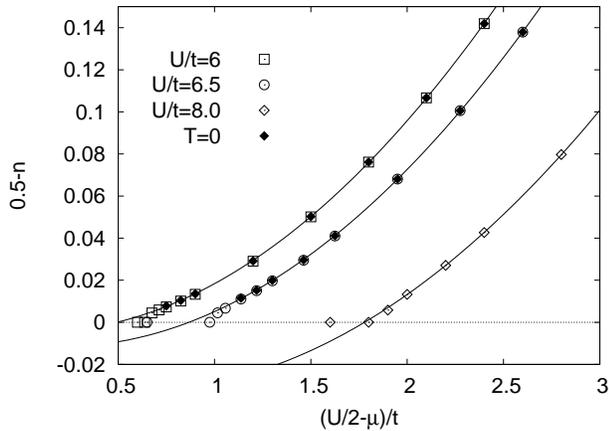}
\caption{Close-up view of the small doping results for $U\gtrsim U_{c2}$, showing the essentially linear onset of doping, which becomes more pronounced as one moves away from the critical point. The lines show parabolic fits to the data points which were extrapolated to $T\rightarrow 0$ ($U/t=6$, 6.5) or can be considered indistinguishable from that limit ($U/t=8$). We assume that these curves correspond to $n(T=0,\mu)$.}
\label{lin_fit}
\end{figure}

\begin{figure}[ht]
\centering
\includegraphics [angle=-90, width= 8.5cm] {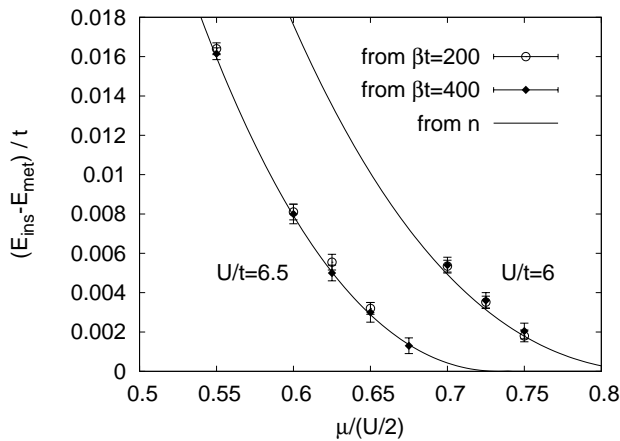}
\caption{Energy difference between the metallic and insulating solutions at $T=0$, extrapolated 
from the data for $\beta t=200$ and $\beta t=400$, respectively. The lines show the result 
obtained from the $n(T=0, \mu)$, assuming a second order transition.
}
\label{OmegaT0}
\end{figure}

\subsection{Second order transition at $T=0$}

\begin{table}[b]
\caption{Location of the $T=0$ second order phase transition ($\mu_{c2}$), compressibility per spin $\partial n/\partial \mu$, and coefficient of the quadratic term $B(\mu-\mu_{c2})^2$ for different values of $U/t$.}
\begin{tabular}{llll}
\\
\hline
$U/t$ \hspace{5mm} & $|\mu_1-\mu_{c2}|/t$\hspace{5mm} & $\partial n/\partial \mu |_{\mu_{c2}} t$\hspace{5mm} & $|B| t^2$\\
\hline
6 & 0.49 & 0.022 & 0.027\\
6.5 & 0.87 & 0.035 & 0.026\\
8 & 1.78 & 0.055 & 0.023\\
12 & 3.90 & 0.075 & 0.018\\
\end{tabular}
\label{tab1}
\end{table}

To address the nature of the transition at $T=0$ we must first extrapolate the measured densities to 
the insulating density $n=0.5$.  
In the range of chemical potentials for which a metallic state is stable for both $\beta=200/t$ and 
$\beta=400/t$ we extrapolate the density to $T=0$ using the Fermi liquid relation $n(T)=n(T=0)+\alpha T^2$ and fitting
$n(T=0)$ and $\alpha$.  Figure~\ref{lin_fit} shows as solid points the result of the extrapolation to $T=0$ and 
as open symbols the computed density at our lowest temperature. One sees that in the density 
range ($0.5-n\gtrsim 0.01$) where more than one temperature is available, the $\beta=400/t$
data are essentially converged to the $T=0$ value. 
%A roughly linear dependence on doping is evident in the region $0.5-n\lesssim 0.03 - 0.05$ (depending on $U$), implying a constant compressibility. 
The roughly linear dependence of doping on $(\mu-\mu_1)$ in the 
region $0.5-n\lesssim 0.03-0.05$ (depending on $U$) implies a constant compressibility. 
Figure~\ref{lin_fit} shows that the density can be fitted very well over the entire measurement
range to the function $n(T=0,\mu)=A(\mu-\mu_{c2})+B(\mu-\mu_{c2})^2$.
Performing the fit, we find the parameters listed in Tab.~\ref{tab1}. 

To verify the consistency of our analysis and determine the order of the transition, we show as lines in Fig.~\ref{OmegaT0} the thermodynamic potential curves obtained by use of our fits to $n(T=0,\mu)$ and the $T=0$ version of Eq.~(\ref{Ffromn}), with $\mu_c$ set equal to the value $\mu_{c2}$ at which $n=0.5$ and the integration constant $\Omega_\text{met}(\mu_c)$ set equal to $E_\text{ins}$,  
\begin{equation}
E_\text{met}(\mu)=E_\text{ins}(\mu_{c2})-2\int_{\mu_{c2}}^\mu d\mu' n(T=0,\mu').
\label{Emet_E_ins}
\end{equation}
These curves are based on the assumption that at $T=0$ the energies of metallic and insulating states coincide only at the chemical potential $\mu=\mu_{c2}$ at which $n=0.5$, and that the metal-insulator transition is hence continuous. 

\begin{figure}[t]
\centering
\includegraphics [angle=-90, width= 8.5cm] {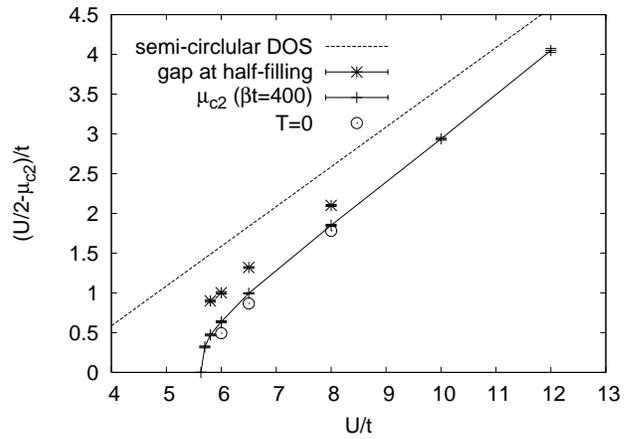}
\caption{Open circles: critical chemical potential $\mu_{c2}$ determined by the extrapolation of the density 
to temperature $T=0$. Crosses: position of the spinodal point $\mu_{c2}(\beta t=400)$ (where the metallic 
solution ceases to exist) as a function of $U$. 
The stars indicate the size of the spectroscopic gap at half-filling, $\Delta$, as determined by 
analytic continuation of insulating Green functions for $\beta t=40$, and the solid line gives a rough 
estimate for this gap, which assumes a semi-circular density of states.
}
\label{mu_c}
\end{figure}

On the other hand, we can extrapolate the measured thermodynamic potentials $\Omega_\text{met}(\mu, T)$ at $\beta=400/t$ and $\beta=200/t$ 
to $T=0$ using the estimated $\gamma$ values and Eq.~(\ref{FofT}). These results are shown in Fig.~\ref{OmegaT0} as solid and open points, respectively.
The close agreement between the two estimates for the energy difference
verifies the analysis and shows that at $T=0$ %the energies of metallic and insulating states coincide only at the chemical potential at which $n=0.5$. We therefore conclude that 
(unlike at $T>0$) the doping does not jump discontinuously
as the chemical potential is increased into the metallic region.

The critical chemical potentials $\mu_{c2}\equiv \mu_{c2}(T=0)$ (measured from the half filling value $\mu_1$) are shown as open circles
in Fig.~\ref{mu_c} while the positions of the spinodal points obtained from our solution
at $\beta=400/t$ are shown as crosses connected by the solid line. 
One sees that the zero temperature extrapolation is important for elucidating the behavior
in the region near $U_{c2}$.  We remark that our $T\rightarrow 0$ extrapolations agree well with the zero temperature results for $\mu_{c2}$ presented very recently in Ref.~\onlinecite{Garcia06}. 

It is interesting to compare the chemical potentials  $\mu_{c2}$ with estimates of the gap
at half filling. The stars in Fig.~\ref{mu_c} indicate the size of the spectroscopic gap $\Delta$
obtained by analytically continuing Monte Carlo data for the higher temperature
$\beta=40/t$.\cite{Armin} We see that the doping-induced states occur much before the chemical potential
reaches the edge of the band; thus doping induces ``in-gap" states. However, a glance
at Fig.~\ref{n} shows that by the time the density is increased beyond a few percent, 
the chemical potential is inside the Hubbard bands. The in-gap nature of the states is therefore relevant
only at extremely low dopings of a few percent or less. 

A scaling behavior is evident for interaction strengths near $U_{c2}$.  In particular
both the compressibility and the critical chemical potential vanish as
$U\rightarrow U_{c2}^+$ but the ratio remains roughly constant (see Tab.~\ref{tab1}). 
A simple scaling analysis would suggest that 
\begin{equation}
n(\mu)-0.5={\bar n} \left(\mu-\mu_1\right)^x\Phi\left(\frac{\mu-\mu_1}{\mu_{c2}-\mu_1}\right)
\label{scaling}
\end{equation}
with $\Phi(y)$ a function tending to a constant as $y\rightarrow \infty$ and vanishing at $y=1$, but with a non-vanishing first derivative.
The quadratic dependence of $n$ on $\mu$ at larger chemical potentials and the rough
scaling of $\partial n/\partial \mu |_{\mu_{c2}}$ and $(\mu_{c2}-\mu_1)$ suggest an exponent $x=2$.
We do not have sufficient accuracy to determine precisely the scaling function and the
behavior of $\mu_{c2}$. These
depend crucially on the value of $U_{c2}$, which we have not determined with
precision. If the value $U_{c2}=5.8t$ quoted in Ref.~\onlinecite{Bulla99} is used, our data are consistent
with the relation
\begin{eqnarray}
\mu_{c2}-\mu_1&=&
C_\mu 
\left(\frac{U-U_{c2}}{t}\right)^{1/2}.
%\left. \frac{\partial n}{\partial \mu}\right|_{\mu_{c2}}&=& \chi \left(\mu_{c2}-\mu_1\right)
\end{eqnarray}

\begin{figure}[t]
\centering
\includegraphics [angle=-90, width= 8.5cm] {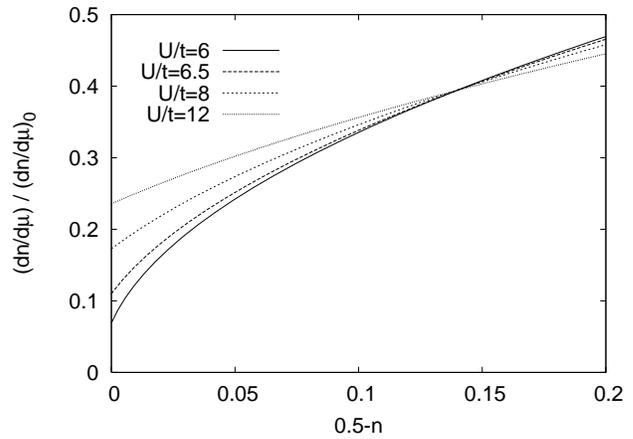}
\caption{$\partial n/\partial \mu$ normalized to the non-interacting value $(\partial n/\partial \mu)_0\approx 1/\pi t$, as a function of density per spin, $0.5-n$. For $U$ close to $U_{c2}$, the compressibility is substantially reduced relative to the non-interacting system.  
}
\label{compressibility}
\end{figure}

\section{Conclusions}

We have studied the doping dependent Mott transition in the one-band Hubbard model, using single site DMFT and a powerful diagrammatic QMC impurity solver which allows access to low temperatures even at strong interactions. 
A detailed quantitative understanding of the doping driven metal-insulator transition could be obtained. By computing the temperature dependence of the energy and thermodynamic potential we were able to perform a convincing extrapolation to $T=0$ and show that while the metal-insulator transition at $T>0$ is first order (with a jump in density), it becomes continuous at $T=0$. At the critical chemical potential $\mu_{c2}(U)$, the density per spin, $n$, smoothly approaches the Mott insulating value 0.5. Our data are consistent with the scaling assumption that $\mu_{c2}(U)$ goes smoothly to the half filled band value $\mu_1=U/2$ as $U\rightarrow U_{c2}^+$. Our results are in substantial agreement with a very recently published density matrix renormalization group study of the same model.\cite{Garcia06} This study determined $\mu_{c2}(U)$ as the boundary of a coexistence region, without making a statement on the order of the transition, finding for example $(\mu_{c2}-\mu_1)(U=6t)\approx 0.5t$ and $(\mu_{c2}-\mu_1)(U=6.5t)\approx 0.9t$, in good agreement with the estimates presented in Tab.~\ref{tab1}.

We determined the behavior of the electronic compressibility $\partial n/\partial \mu$ as a function of $U$ and doping, finding that it vanishes at $U=U_{c2}$, $\mu=\mu_1$ and grows roughly linearly with distance in $U$ and $\mu$ from this critical point. The vanishing of the compressibility at the $T>0$ critical end point of the Mott transition has been extensively discussed in the literature.\cite{compressibilityref, Fournier03}  
In a series of interesting publications, Imada and co-workers have argued that
at the density-driven $T=0$ metal-insulator transition the compressibility
$\partial n/\partial \mu$ should vanish,\cite{Imada98comment} in contrast to our finding that the quantity is non-vanishing for $U>U_{c2}$. The conclusions of Imada and co-workers are based on  hyperscaling, which is unlikely to apply in the limit of spatial dimensionality $d \rightarrow \infty$ in which the DMFT approach is exact. Further consideration of this issue in finite dimensionality is an important open problem.

The values we obtain for the electronic compressibility are interesting. 
Figure~\ref{compressibility} shows the compressibility, normalized to the non-interacting value of approximately $1/\pi t$, as a function of doping. These curves were obtained from the fitting functions for $n(T=0,\mu)$ and show that the suppression of the compressibility at $\mu=\mu_1$ and $U=U_{c2}$ persists over a wide interaction and doping range. This suppression has two experimental consequences: first, the square of the inverse Thomas-Fermi screening length
\begin{equation}
q_\text{TF}^2=\frac{4\pi e^2}{\epsilon}\frac{\partial n}{\partial \mu}
\end{equation} 
should be strongly reduced %enhanced 
near the metal-insulator transition, possibly leading to unusually weak screening of charged impurities. However, a simple estimate for high-$T_c$ materials gives $\partial n/\partial \mu \approx 1/eV$. A lattice constant of $4\text{\AA}$ and a background dielectric constant of 10 would then imply $q_\text{TF}^2|_\text{band}\approx 4\text{\AA}^{-2}$, so even the largest renormalization shown in Fig.~\ref{compressibility} would only lead to a $q_\text{TF}\approx 1 \text{\AA}^{-1}$. 
Thus screening is always expected to be at the scale of a lattice constant. Nevertheless the effect might be observable in scanning tunneling microscopy. 
Possibly more easily observable would be a doping dependence of the sound velocity via the Bohm-Staver relation $c^2\sim \partial n/\partial \mu$.\cite{Fournier03}

Finally, looking towards the future, we suggest that using the techniques presented here and in Ref.~\onlinecite{Werner06} to reexamine the metal-insulator transition in the context of cluster dynamical mean field theories is an urgent open problem.

\acknowledgements

Support from NSF DMR 0431350 is gratefully acknowledged. We thank A. Comanac for computing the band gaps in Fig.~\ref{mu_c} by means of analytical continuation. The calculations have been performed on the Hreidar Beowulf cluster at ETH Z\"urich, using the ALPS library.\cite{ALPS} We thank M. Troyer for the generous allocation of computer time.

\end{document}